# Analysis of quantum decay law: Is quantum tunneling really exponential?

M. S. Hosseini-Ghalehni, B. Azadegan[a], S. A. Alavi[b]

*Department of Physics, Hakim Sabzevari University, P.O. Box 397, Sabzevar, Iran.*

**Abstract:** The exponential decay law is well established since its first derivation in 1928; however, it is not exact but only an approximate description. In recent years some experimental and theoretical indications for non-exponential decay have been documented. First we solve analytically the time-dependent Schrödinger equation in one dimension for a potential consisting of an infinite wall plus a rectangular barrier with finite width and also a cut harmonic oscillator potential by considering it as a sequence of square potentials. Then using the staggered Leap-Frog method, we solve the time-dependent Schrödinger equation for the cut harmonic oscillator potential. In both methods, time dependence of the survival probability of the particle and the decay parameter $\lambda$ are analyzed. The results exhibit non-exponential behavior for survival probability at short and intermediate times.

1. **Introduction**

Tunneling through potential barriers that classically cannot be overcome is at the heart of quantum mechanics and plays an essential role in fundamental dynamical phenomena like Josephson oscillations, tunnel diode quantum computing and the scanning tunneling microscope. It turns up in the description of disordered one-dimensional lattices and realistic one-dimensional solid-state systems, such as quantum wells, junctions and super lattices. It has a significant function in many branches of science from alpha decay, fusion and fission in nuclear physics to photo association and photo dissociation in biology and chemistry. This problem also plays a key role in the study of electronic transport in polymers and the effects of intermolecular potential barriers in chemical reactions. Moreover, tunneling is one of the major mechanisms that stimulate the loss of particles from quantum systems which occurs whenever a quantum system is coupled via a potential barrier to a continuum, i.e., to a set of asymptotically free states. Such particle loss is widely encountered in nature, e.g., in radioactive decay or in the auto ionization of excited atomic states. In nuclear physics many types of heavy atomic nucleus spontaneously decay to produce daughter nuclei via the emission of particles of some characteristic energy. Nuclear fusion occurs in main sequence stars like the Sun. The tunneling process is also of crucial importance in many body systems.

[a] e-mail: azadegan@hsu.ac.ir

[b] e-mails: s.alavi@hsu.ac.ir; alaviag@gmail.com (corresponding author)

It is also a well-known fact that decay of unstable states is described by the exponential decay law. The exponential law in alpha decay was first explained by Gamow in 1928 [1]. The decay law of unstable systems plays a crucial role in different area of physics: the electromagnetic decays of atoms, the decays of radioactive nuclei, decays of hadronic resonances like top quarks and Higgs bosons and the Standard Model particles. Most of the unstable physical systems such as electromagnetic decay of atoms and nuclei, beta and alpha decay of radioactive nuclei are all described by the exponential decay law. Understanding the decay dynamics of unstable isolated systems is of relevance to a wide variety of fields ranging from quantum science to statistical mechanics and cosmology.

Although the exponential decay law provides a good description for quasi-stationary states and decay phenomena, it is only an approximate solution. Recently some experimental and theoretical evidences for non-exponential decay have been reported. Non-exponential decays and relaxations have been observed and studied extensively in physics, from the viewpoint of condensed matter to atomic and nuclear physics, see e.g., [2–36]. A recent and really challengeable observation of non-exponential decay in nuclear physics is GSI anomaly. That is the periodic modulation of the expected exponential law in the EC-decays of different highly charged ions, stored at GSI, observed by the FRS/ESR Collaboration [19,20]. Many attempts have been made to explain this observation since 2008, which some of them are provided in the references [21-35]. It is worth mentioning that, the relation between the quantum mechanical survival probability of an unstable system in motion and that of the system at rest has been investigated in [37].

The structure of the paper is as follows: In section 2, we solve analytically the time-dependent Schrödinger equation for a potential consisting of an infinite wall plus a rectangular barrier with finite width. Analytical solutions of the Schrödinger equation for a cut harmonic oscillator potential are presented in Section 3. The same problem is studied using Staggered Leap-Frog Method, in section 4. Finally the effects of increasing the width of the cut harmonic oscillator barrier on the survival probability are investigated in section 5.

**2. Analytical solution of time-dependent Schrödinger equation with a potential consisting of an infinite wall plus a rectangular barrier with finite width**

The general form of the one-dimensional Schrödinger equation is

$$i\hbar \frac{\partial \psi(x,t)}{\partial t} = -\frac{\hbar^2}{2m}\frac{\partial^2 \psi(x,t)}{\partial t^2} + v(x,t)\psi(x,t). \tag{1}$$

We set $\hbar = m = 1$, for the numerical reasons. In this case, the time-independent one-dimensional Schrödinger equation takes the following form:

$$-\frac{1}{2}\frac{d^2\psi_E(x)}{dx^2} + v(x)\psi_E(x) = E\psi_E(x) \tag{2}$$

We define the one-dimensional rectangular barrier as follows:

$$v(x) = \begin{cases} \infty & x \leq 0 \\ 0 & 0 < x < a \\ v_0 & a \leq x \leq b \\ 0 & x > b \end{cases} \tag{3}$$

with $a, b > 0$ and $v_0 > 0$. At $x = 0$, $V$ is assumed to be infinite, so $\psi(x = 0, t) = 0$. For $x=a$ and $x=b$, we have the following boundary conditions:

$$\psi_E^I(a) = \psi_E^{II}(a), \quad \frac{d\psi_E^I(x)}{dx}\bigg|_{x=a} = \frac{d\psi_E^{II}(x)}{dx}\bigg|_{x=a} \tag{4}$$

$$\psi_E^{II}(b) = \psi_E^{III}(b), \quad \frac{d\psi_E^{II}(x)}{dx}\bigg|_{x=b} = \frac{d\psi_E^{III}(x)}{dx}\bigg|_{x=b} \tag{5}$$

The solution of the time-dependent Schrödinger equation (1) could be written as the energy-convolution integral,

$$\psi(x,t) = \int_0^\infty \phi(E)\psi_E(x)e^{-iEt}dE \tag{6}$$

where $\phi(E)$ is a spectral function such that the integral is convergent for all values of *x*, *t* and the resulting wave function is square-integrable. Note that square-integrability of $\psi(x,t)$ also requires $E$ to be real. The overall normalization constant is, then, calculated from

$$\int_0^\infty \psi^*(x,t)\psi(x,t)dx = 1 \tag{7}$$

The solutions of the time-independent Schrödinger equation (2), for $E \geq 0$ are

$$\psi_E^I(x) = A\,Sin(k_1 x), \qquad 0 \leq x < a \tag{8}$$

$$\psi_E^{II}(x) = B_1 e^{k_2 x} + B_2 e^{-k_2 x}, \qquad a < x < b \tag{9}$$

$$\psi_E^{III}(x) = C_1 Cos[k_1 x] + C_2 Sin[k_1 x], \qquad x > b \tag{10}$$

where $k_1 = \sqrt{2E_n}$, $k_2 = \sqrt{-2(E_n - v_0)}$ and $A, B_{1,2}, C_{1,2}$ are constants. These constants are obtained from boundary conditions (4) and (5). By selecting $A$ as the overall normalization constant, the boundary conditions yield:

$$B_1 = \frac{e^{-ak_2}(k_1 \cos[ak_1] + k_2 \sin[ak_1])A}{2k_2} \tag{11}$$

$$B_2 = \frac{e^{ak_2}(-k_1 \cos[ak_1] + k_2 \sin[ak_1])A}{2k_2} \tag{12}$$

$$C_1 = \frac{A}{2k_1 k_2} e^{-(a+b)k_2} \begin{pmatrix} e^{2bk_2}(k_1 \cos[ak_1] + k_2 \sin[ak_1])(k_1 \cos[bk_1] - k_2 \sin[bk_1]) + \\ e^{2ak_2}(-k_1 \cos[ak_1] + k_2 \sin[ak_1])(k_1 \cos[bk_1] + k_2 \sin[bk_1]) \end{pmatrix} \tag{13}$$

$$C_2 = \frac{A}{2k_1 k_2} e^{-(a+b)k_2} \cos[bk_1] \begin{pmatrix} -e^{2ak_2}(-k_1 \cos[ak_1] + k_2 \sin[ak_1])(k_2 - k_1 \tan[bk_1]) + \\ e^{2bk_2}(k_1 \cos[ak_1] + k_2 \sin[ak_1])(k_2 + k_1 \tan[bk_1]) \end{pmatrix}. \tag{14}$$

The spectral function is calculated as:

$$\phi(E_n) = A^2/(C_1^* C_1 + C_2^* C_2) \tag{15}$$

The overall normalization constant is finally calculated from (7). Fig. 1 shows the wave functions versus $x$ for three different energies E= 1.5, 2.5, 3.0. Fig. 2 illustrates the spectral function which is calculated using Eq. (15) for the infinite wall plus repulsive rectangular potential function for a=1.5, b=2.25 and $v_0$=3.

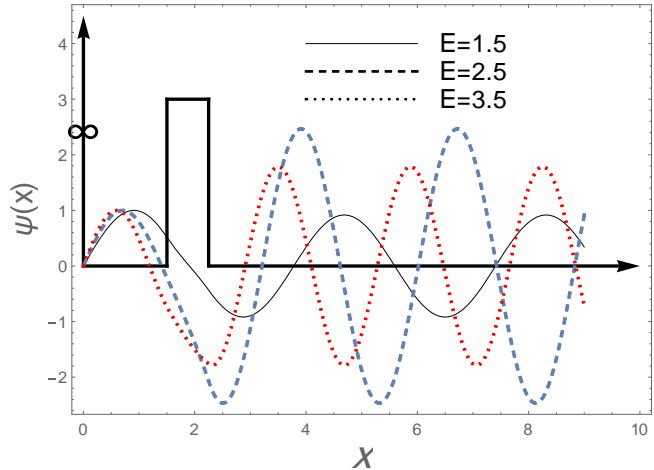

Fig.1. The wave functions for a potential consisting of an infinite wall and a repulsive rectangular potential for three different energies E= 1.5, 2.5, 3.0. In this plot a=1.5, b=2.25 and $v_0$=3.

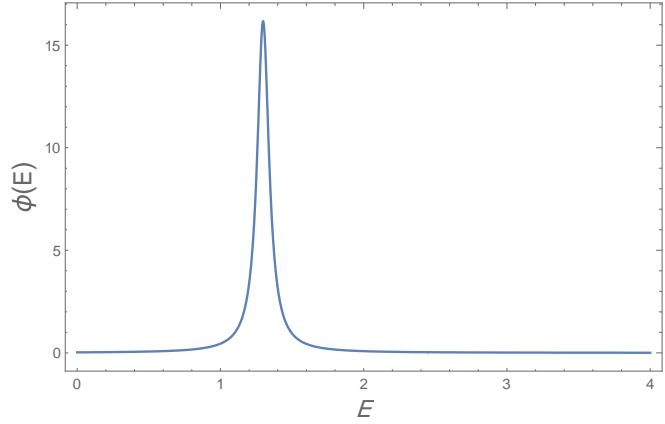

Fig.2. Spectral function for the infinite wall plus repulsive rectangular potential function for a=1.5, b=2.25 and v₀=3.

The survival probability is, then, defined as:

$$P_{in}(t) = \int_0^a \psi^*(x,t)\,\psi(x,t)dx \tag{16}$$

which will be calculated numerically. In order to investigate the effects of increasing the width of the barrier on the survival probability and the decay parameter, we have solved the time-independent Schrödinger equation in one dimension for different widths $b - a = 0.2, 0.75, 1.0,$ and $1.5$.

Logarithmic plots of the survival probabilities (the lack of tunneling through potential barrier) for an infinite wall plus repulsive rectangular potential with different widths versus time are shown in Fig. 3. It can be seen that by increasing the width of the rectangular potential the survival probabilities decreases. The logarithm of the survival probabilities for widths 0.75, 1.0, and 1.5 tend to the straight lines. This behavior is consistent with WKB approximation. For a width with value 0.2, the logarithm of survival probability deviates from the straight line, which means that, the exponential decay law is not valid for small widths. One successful experimental observation of violation of the exponential decay law was made by Rothe et.al., [38], in measurement of the luminescent decay of dissolved organic materials.

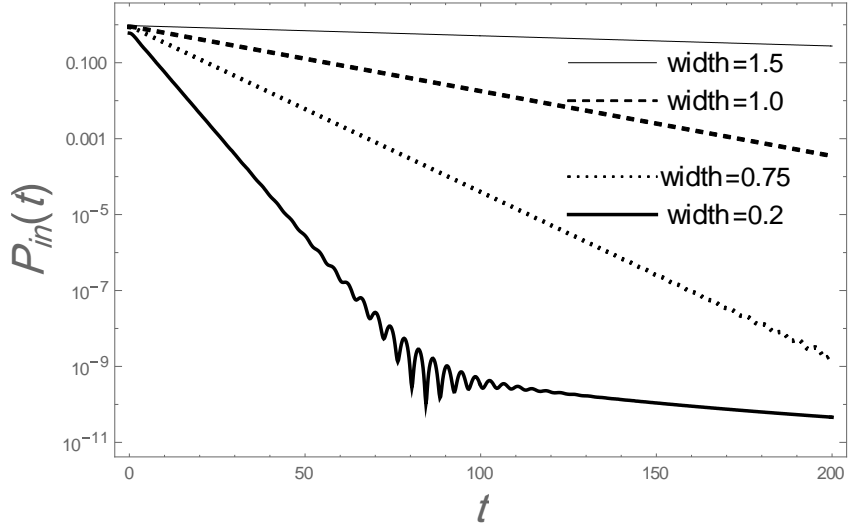

Fig.3. Logarithmic plot of the survival probability for a potential consisting of an infinite wall and a repulsive rectangular potential versus time. In this plot a=1.5, v₀=3 and $b - a = 0.2, 0.75, 1.0,$ and $1.5$.

The decay parameter λ could be defined by:

$$\lambda(t) = -\frac{1}{P_{in}}\frac{dP_{in}}{dt} = -\frac{d}{dt}[ln(P_{in}(t))] \qquad (17)$$

The decay parameters for widths $b - a = 0.75, 1.0$ and $1.5$ are plotted versus time in Fig. 4. They increase very fast in time and reach to their maximal values $\lambda_{max}$. It could be seen that by increasing the width of the rectangular potential the decay parameter decreases and the amplitude of its oscillation decreases as well. The oscillation in the amplitude is obviously observed for smaller widths.

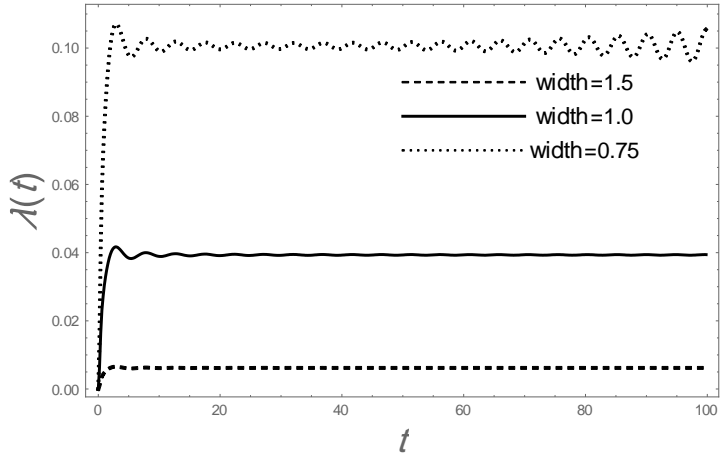

Fig.4. The decay parameter $\lambda$ for a potential consisting of an infinite wall and a rectangular potential with $width = b - a = 0.75, 1.0,$ and $1.5$. The oscillations in the amplitudes is obviously observed for smaller widths.

It is important to note that, decay is exponential when the integration interval is from negative infinity to infinity. Considering the finite interval for the energy states i.e., having a cut-off leads to deviation from the exponential decay [24]. In other words, the non-exponential behavior is related to the cut-off in the energy interval. If the energy is to vary over the entire real axis, then the residue theorem will yield exponential decay [36]. The integral in Eq. (6) is from zero to infinity not from minus infinity to plus infinity. It is because, the peak of the spectral function in Fig. 2 is located in the positive part of the horizontal axis, so only the positive energies are contributed in the problem (integral).

## 3. Numerical solution of the Schrödinger equation for a cut harmonic oscillator potential

It is well-known in quantum mechanics that, the Schrödinger equation could be solved for a potential barrier with an arbitrary shape, by approximating it with a juxtaposition of square potential barriers as shown in Fig. 5. To generalize our calculations in previous section to more than one rectangular barrier, we consider a cut harmonic oscillator potential as follows:

$$V(x) = \begin{cases} \frac{1}{2}\alpha x^2 & 0 \leq x \leq \frac{\beta}{2} \\ 0 & other \end{cases} \quad (18)$$

where we take $\alpha = 0.28$ and $\beta = 6.0$.

Now we approximate this potential by one, two, three, four, five and six square barriers and for each one, solutions of the time-independent Schrödinger equation are obtained by applying the boundary conditions of the problem i.e., Eqs. (4) and (5). Due to the large amount of the wave functions coefficients, we have not presented them in the manuscript. The approximation of the cut harmonic oscillator potential with various numbers of square barriers along with the corresponding wave functions for energies 0.08, 0.4 and 1.4, are shown in Fig 6.

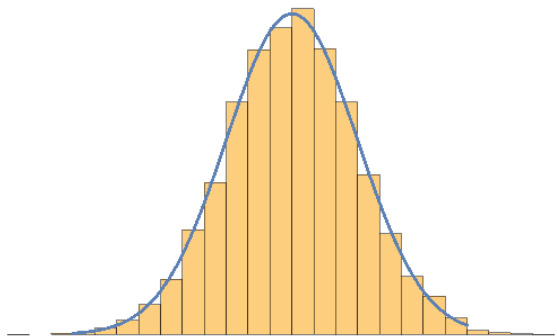

Fig. 5. Approximation of smooth barrier potential by juxtaposition of square potential barriers.

The survival probabilities are calculated numerically using Eq. (16) and presented in Fig. 7. The decay parameters $\lambda$ are calculated using Eq. (17) and are plotted in Fig. 8. It is observed that, the decay parameter $\lambda$ is not constant and has an oscillatory behavior, so the tunneling (decay probability) is not exponential. Furthermore, as we increase the number of barriers, the decay parameters converge to the same value which means that, more increase in the number of barriers does not have significantly more accurate result, so the 5 or 6 square barriers lead to sufficiently accurate results.

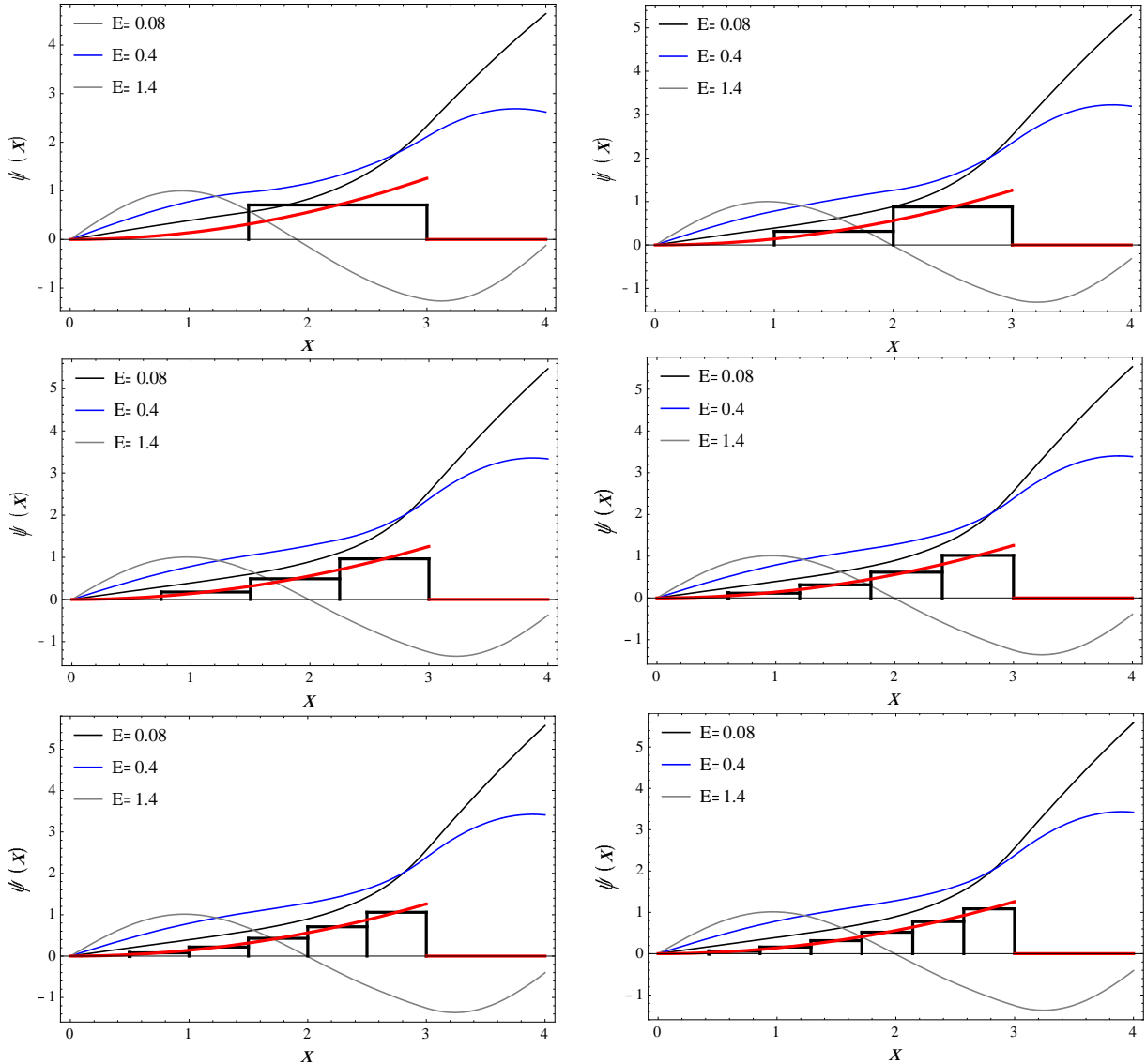

Fig.6. Wave functions for a cut harmonic oscillator potential approximated by one up to six rectangular barriers.

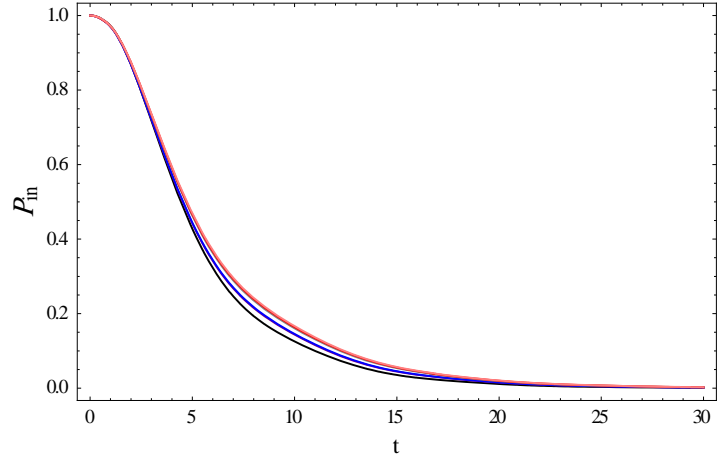

Fig.7. The survival probabilities for a cut harmonic oscillator potential approximated by one (Black), two (Purple), three (Blue), four (Gray), five (Red), and six rectangular barrier potentials (Pink), respectively.

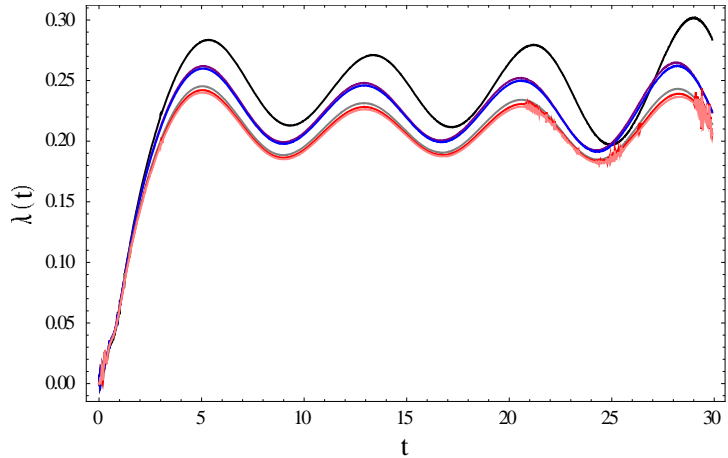

Fig.8. The decay parameter $\lambda$ for approximating a cut harmonic oscillator potential by one to six rectangular barrier potentials.

## 4. Staggered Leap-Frog Method

The second method to solve the time-dependent Schrödinger equation is Staggered Leap-Frog method. In this section we apply this method to the cut harmonic oscillator potential.

It is noteworthy to mention, we have used as an initial condition a localized Gaussian distribution wave packet of width σ centered at $x=x_0$. This initial condition does not exactly satisfy the boundary conditions, but it is very close to the exact one. We also use boundary condition as, $\psi(x,t) = 0$ at $x = 0$ and infinity.

The initial probability density, the probability density at time 30 and the survival probability for a cut harmonic oscillator potential computed using Staggered Leap-Frog method are presented in Figs. 9-11 respectively. The decay parameter $\lambda$ is avalable in Fig. 12, which shows that, it is not constant so the decay is not exponential. The results of Staggered Leap-Frog method are in good agreement with the results of our first method.

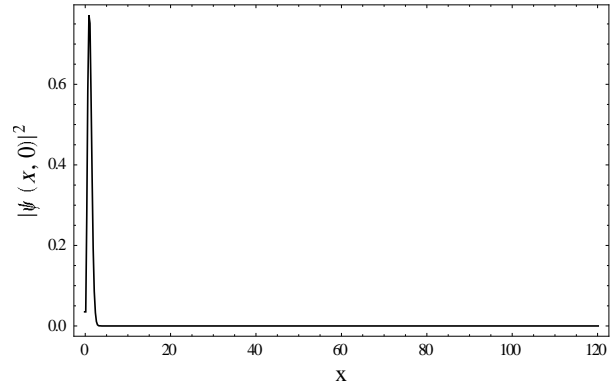

Fig.9. The initial probability density for a cut harmonic oscillator potential.

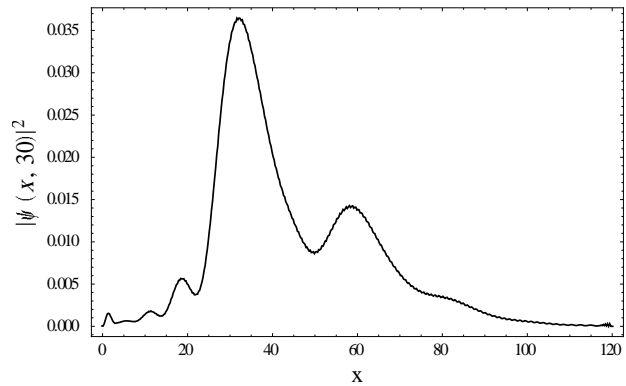

Fig.10. The probability density at $t = 30$ for a cut harmonic oscillator potential employing Staggered Leap-Frog method.

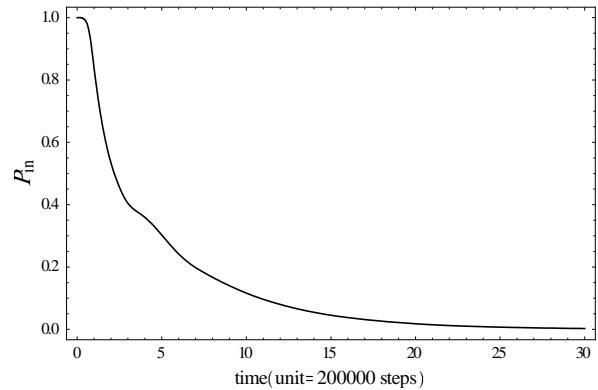

Fig.11. The survival probabilities for a cut harmonic oscillator potential using Staggered Leap-Frog method.

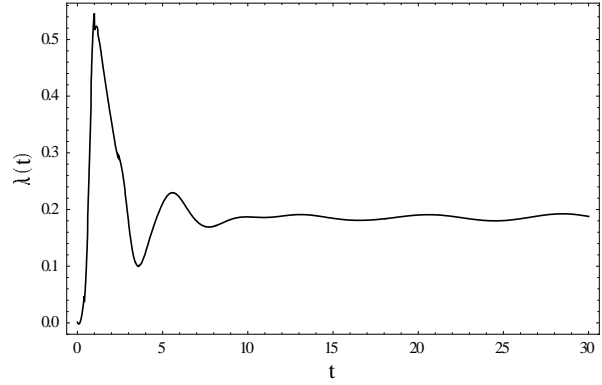

Fig.12. The decay parameter $\lambda$ for a cut harmonic oscillator potential computed by Staggered Leap-Frog method.

## 5. The effects of increasing the width of the cut harmonic oscillator barrier

In order to investigate the effects of increasing the width of the barrier on the survival probability and on the decay parameter, we employ again the two methods we have used so far, as follows: a). We approximate the cut harmonic oscillator barrier by six square barriers and solve analytically the time-dependent Schrödinger equation. b). We Use the Staggered Leap-Frog method. In both cases, we increase the width of the barrier as $\beta = 5, 5.5$ and $6.0$ and compare the results.

The survival probabilities and the decay parameters for methods (a) and (b) are shown in Figs. 13-16, respectively. As it is observed, $P_{in}$ falls faster to zero as the width of the barrier decreases. It is also shown that the decay parameter $\lambda$ is not constant for any value of $\beta$, so the tunneling(decay) is not exponential. Furthermore, as we increase the values of $\beta$, the amplitude of oscillations of the decay parameter and the average of the decay parameter decrease. Also as the value of $\beta$ is increased the amplitude of oscillations of the decay parameter tend to a common constant value, which means that as the width of barrier is increased, the decay turns to exponential form for intermediate times.

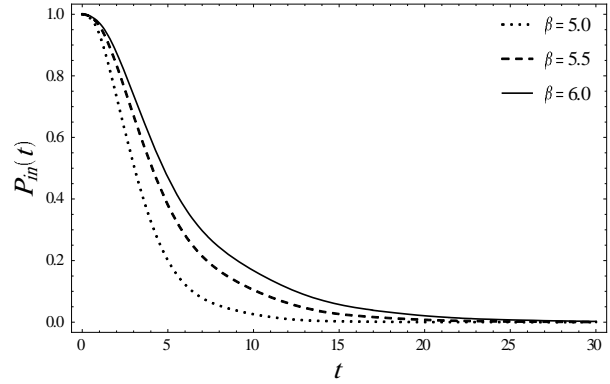

Fig.13. The survival probabilities computed using approximating a cut harmonic oscillator potential by six rectangular potential barriers, for $\beta = 5.0, 5.5$ and, $6.0$.

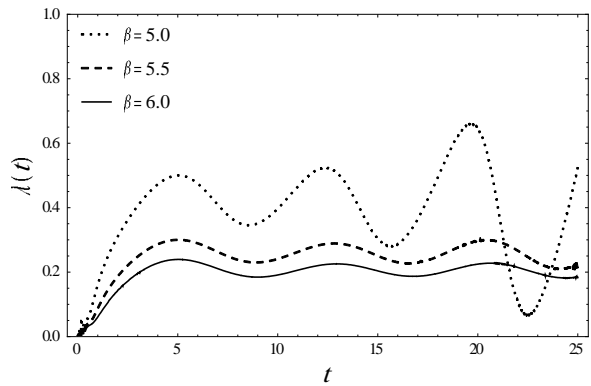

Fig.14. The decay parameter $\lambda$, obtained using approximating a cut harmonic oscillator potential by six rectangular potential barriers, for $\beta = 5.0, 5.5$ and, $6.0$.

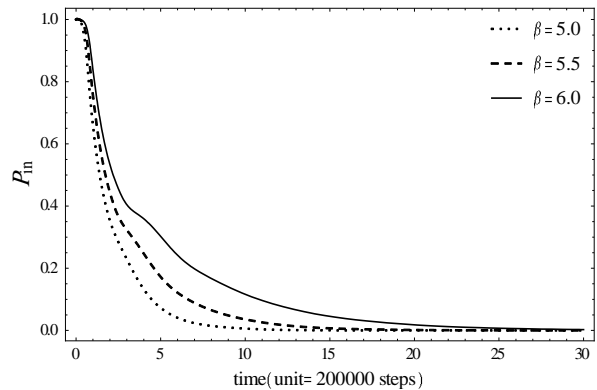

Fig.15. The survival probabilities for a cut harmonic oscillator potential with $\beta = 5.0, 5.5$ and, $6.0$, making use of Staggered Leap-Frog method.

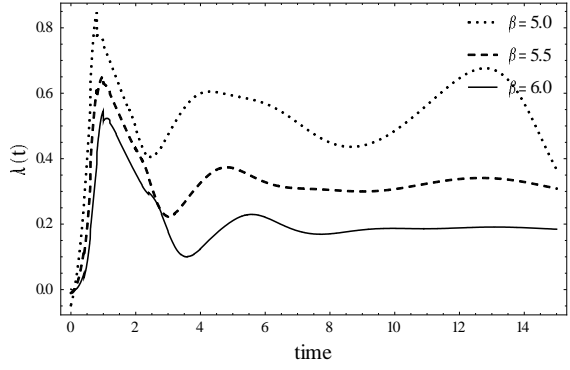

Fig.16. The decay parameters for a cut harmonic oscillator potential with $\beta = 5.0, 5.5$ and, $6.0$, obtained by Staggered Leap-Frog method.

## 6. DISCUSSION

Let $|\alpha\rangle$ be an unstable state at $t = 0$. For $t > 0$, the amplitude of survival probability is given by:

$$a(t) = \langle \alpha | e^{-iHt} | S \rangle, \quad \hbar = 1 \tag{19}$$

So, for the survival probability we have $P(t) = |a(t)|^2$. The Taylor expansion of the amplitude is as follows:

$$a(t) = 1 - i t \langle \alpha | H | \alpha \rangle - \frac{t^2}{2} \langle \alpha | H^2 | \alpha \rangle + \cdots \tag{20}$$

So:

$$P(t) = |a(t)|^2 = a^*(t) \, a(t) = 1 - \frac{t^2}{t_z^2} + \cdots \tag{21}$$

where $t_z$ is the Zeno time and $t_z^{-2} = \langle \alpha | H^2 | \alpha \rangle - \langle \alpha | H | \alpha \rangle^2$. The total Hamiltonian $H$ can be decomposed into two parts [39]: $H = H_0 + H_{int}$, $H_0 = PHP + QHQ$, $H_{int} = PHQ + QHP$ where $P = |\alpha\rangle\langle\alpha|$ and $Q = 1 - P$.

The initial state is an eigenstate of the free Hamiltonian $H_0$, which is off-diagonal with respect to the interaction Hamiltonian:

$$H_0 |\alpha\rangle = \alpha |\alpha\rangle, \quad \langle \alpha | H_{int} | \alpha \rangle = 0 \tag{22}$$

So the Zeno time depends only on the square of the interaction Hamiltonian:

$$t_z^{-2} = \langle \alpha | H_{int}^2 | \alpha \rangle \tag{23}$$

In order to observe a delay of the tunneling time i.e. a decrease of the decay constant $\lambda$ as a consequence of frequent consecutive measurements, the time interval $\Delta t$ between two consecutive measurements should be smaller compared to the Zeno time $t_z$. Now, let us estimate the Zeno time in our study. We consider a constant rectangular barrier with the value in $MeV$ range e.g., $4MeV$, so from Eq. (23), we have:

$$t_z = 1.7 \times 10^{-22} \, s \tag{24}$$

For a barrier with the value $4eV$ we get the following result:

$$t_z = 1.7 \times 10^{-16} \, s \tag{25}$$

The values of Zeno times are too short (even unreal), so the influence of frequent consecutive measurements on the value of decay parameter $\lambda$ could be neglected. It should be mentioned, although the influence of Zeno effect on the value of $\lambda$ could be neglected, but it may has influence on the amplitude of time modulation of $\lambda$. It is also worth noting that, in [26], it is shown that the influence of Zeno effect on the decay constant $\lambda_{EC}$ of GSI anomaly can be neglected.

## 7. CONCLUSION

Analytical solution of time-dependent Schrödinger equation with a potential consisting of an infinite wall plus a rectangular barrier with finite width, is presented. The probability for the particle to be found inside the potential well as a function of time is calculated. Then numerical solution of the Schrödinger equation for a cut harmonic oscillator potential is done. The solutions of the time-dependent Schrödinger equation with a cut harmonic potential using the staggered Leap-Frog method are also obtained. In both approaches, the decay parameter for short times increases from zero to a maximum, and for intermediate times has an oscillatory behavior around a constant value. Furthermore, for both methods it was shown that with increasing the width of the barrier the amplitude of oscillations of the decay parameter decrease.

## ACKNOWLEDGEMENTS

We would like to express our gratitude to Francesco Giacosa for his comments to improve this manuscript. S. A. Alavi is highly delighted and thankful of INFN, Turin and specially Carlo Giunti. It was there that the initial spark of this work was ignited.

**References**

[1] G. Gamow, "Quantum Theory of the Atomic Nucleus", Zur Quantentheorie des Atomkernes, ZP **51**, 204 (1928).

[2] G. Andersson et.al., "Non-exponential decay of a giant artificial atom", Nature Physics **15**, 1123 (2019).

[3] J. Kumlin, et.al., "Nonexponential decay of a collective excitation in an atomic ensemble coupled to a one-dimensional waveguide", Phys. Rev. A **102**, 63703 (2020).

[4] S. R. Wilkinson, et.al., "Experimental evidence for non-exponential decay in quantum tunneling", Nature **387**, 575 (1997).

[5] M. Peshkin, A. Volya, and V. Zelevinsky, "Non-exponential and oscillatory decays in quantum mechanics", EPL **107**, 40001 (2014).

[6] S. Duki, H. Mathur, "Nonexponential tunneling and control of microwave absorption lineshapes via Fano resonance for electrons on helium", Phys. Rev. B **90**, 1 (2014).

[7] G. García-Calderón, R. Romo, "Nonexponential tunneling decay of a single ultracold atom", Phys. Rev. A **93**(1), 022118 (2016).

[8] Elihu Abrahams, "Nonexponential relaxation and hierarchically constrained dynamics in a protein", Phys. Rev. E **71**(1), 051901 (2005).

[9] Stefano Longhi, "Nonexponential Decay Via Tunneling in Tight-Binding Lattices and the Optical Zeno Effect", Phys. Rev. Lett. **97**, 10 (2006).

[10] F. H. L. Koppens, et.al., "Universal Phase Shift and Nonexponential Decay of Driven Single-Spin Oscillations", Phys. Rev. Lett. **99**, 1 (2007).

[11] G. Iori, E. Marinari and G. Parisi, "Non-Exponential Relaxation Time Scales in Disordered Systems: An Application to Protein Dynamics", EPL (Europhysics Letters) **25**, 491 (1994).

[12] P. J. Aston, "Is radioactive decay really exponential?", EPL (Europhysics Letters) **97**, 52001 (2012).

[13] N. G. Kelkar, M. Nowakowski, and K. P. Khemchandani, "Hidden evidence of nonexponential nuclear decay", Phys.Rev. C Nucl. Phys. **70**, 2 (2004).

[14] Fierro, et.al., "Percolation transition and the onset of nonexponential relaxation in fully frustrated models", Phys. Rev. E **59**, 60 (1999).

[15] A. Wyrzykowski, "Analysis of the Breakdown of Exponential Decays of Resonances", Acta Phys. Pol. B **51**, 2015 (2020).

[16] Zhi-De Chen, Shun-Qing Shen, "Nonexponential relaxation and quantum tunnel splitting in the molecular magnet Fe8", Phys. Rev. B **67**, 012408 (2003).

[17] Q. Guan, et.al., "Nonexponential Tunneling due to Mean-Field-Induced Swallowtails", Phys. Rev. Lett. **125**, 213401 (2020).

[18] A. M. Ishkhanyan, V P Krainov, "Non-exponential tunneling ionization of atoms by an intense laser field" Laser Phys. Lett. **12**, 046002 (2015).

[19] Y. A. Litvinov et al., "Observation of non-exponential orbital electron capture decays of hydrogen-like $^{140}$Pr and $^{142}$Pm ions", Phys. Lett. B Nucl. Elem. Part. High Energy Phys. **664**, 162 (2008).

[20] P. Kienle et al., "High-resolution measurement of the time-modulated orbital electron capture and of the β+ decay of hydrogen-like 142Pm60+ ions", Phys. Lett. B Nucl. Elem. Part. High Energy Phys. **726**, 638 (2013).

[21] F. Giacosa, "Non-exponential decay in quantum field theory and in quantum mechanics: the case of two (or more) decay channels", Foundations of Physics **42**, 1262 (2012).

[22] G. Lambiase, G. Papini, G. Scarpetta, "GSI anomaly and spin–rotation coupling", Phys. Lett. Sect. B Nucl. Elem. Part. High Energy Phys. **718**, 998 (2013).

[23] F. Giacosa, "Energy uncertainty of the final state of a decay process", Phys. Rev. A At. Mol. Opt. Phys. **88**, 1 (2013).

[24]. F. Giacosa and G. Pagliara, "(Oscillating) non-exponential decays of unstable states", PoS BORMIO2012, 28 (2012), arXiv:1204.1896 [nucl-th].

[25]. F. Giacosa and G. Pagliara, "Oscillations in the decay law: A possible quantum mechanical explanation of the anomaly in the experiment at the GSI facility", Quant. Matt. **2**, 54 (2013), arXiv:1110.1669 [nucl-th].

[26]. A. N. Ivanov and P. Kienle, "GSI Oscillations as Laboratory for Testing of New Physics", (2013), arXiv.1312.5206.

[27]. T. Koide, F. M. Toyama, "Decay process accelerated by tunneling in its very early stage", Phys. Rev. A **66**, 064102 (2002).

[28] G. Lambiase, G. Papini, G. Scarpetta, "The role of spin–rotation coupling in the non-exponential decay of hydrogen-like heavy ions", Ann. of Phys. **332**, 143 (2012).

[29] F. C. Ozturk et al., "New test of modulated electron capture decay of hydrogen-like 142Pm ions: Precision measurement of purely exponential decay", Phys. Lett. B **797**, 134800 (2019).

[30] A Gal, "Neutrino signals in electron-capture storage-ring experiments ", *Symmetry 8* (2016).

[31] V. P. Krainov, "Mechanism of "GSI oscillations" in electron capture by highly charged hydrogen-like atomic ions", Sov. J. Exp. Theor. Phys. **115**, 68 (2012).

[32] M. Peshkin, "Ocillating decay rate in electron capture and the neutrino mass difference", Phys. Rev. C **91,** 042501 (2015).

[33] S.-J. Rong, Q.-Y. Liu, " Flavor conservation, micro-causality and GSI anomaly", Modern Phys. Lett. A **27**, 1250093 (2012).

[34] F. Giraldi, "Regularities in the transformation of the oscillating decay rate in moving unstable quantum systems", Eur. Phys. J. D **73**, (2019).

[35] M.S. Hosseini, S. A. Alavi, "Breit–Wigner distribution, quantum beats and GSI Anomaly", Ann. Phys. **410**, 167936 (2019).

[36] A. N. Petridis, et.al., "Exact Analytical and Numerical Solutions to the Time-Dependent Schrödinger Equation for a One-Dimensional Potential Exhibiting Non-Exponential Decay at All Times", J. Modern Phys. **2010**, 124 (2010).

[37] S. A. Alavi, C. Giunti, "Which is the quantum decay law of relativistic particles?", EPL (Europhys. Lett.) **109**, 60001 (2015).

[38] C. Rothe, S.I. Hintschich, A.P. Monkman, "Violation of the Exponential Decay Law at Long Times", Phys. Rev. Lett. **96**, 163601 (2006).

[39] P. Facchi, S. Pascazio, "Quantum Zeno dynamics: mathematical and physical aspects", J. Phys. A Math. Theor. **41**, 493001 (2008).